\documentclass[a4paper]{article}
\usepackage{Odyssey2024}
\usepackage{epsfig,amssymb,amsmath}
\usepackage{graphicx}

\usepackage{amsfonts}

\usepackage{xcolor}
\usepackage{diagbox}
\usepackage{multirow}
\usepackage{booktabs}
\usepackage{subfigure}
\usepackage{mathtools}
\usepackage{amsmath}
\usepackage{cite}
\usepackage{hyperref}
\usepackage{bbm}
\usepackage{algorithm}
\usepackage{algorithmic}
\usepackage[marginal]{footmisc}
\usepackage{bm}
\newcommand\blfootnote[1]{%
  \begingroup
  \renewcommand\thefootnote{}\footnote{#1}%
  \addtocounter{footnote}{-1}%
  \endgroup
}
\ninept

\title{Converting Anyone's Voice: \\ End-to-End Expressive Voice Conversion with a Conditional Diffusion Model}

\name{Zongyang Du$^1$, Junchen Lu$^2$, Kun Zhou$^2$, Lakshmish Kaushik$^3$, Berrak Sisman$^1$}

\address{$^1$ Speech \& Machine Learning (SML) Lab, The University of Texas at Dallas, USA \\
$^2$  Department of Electrical and Computer Engineering, National University of Singapore\\
$^3$ Sony Interactive Entertainment, USA}
\begin{document}
\maketitle

\begin{abstract}
Expressive voice conversion (VC) conducts speaker identity conversion for emotional speakers by jointly converting speaker identity and emotional style. Emotional style modeling for arbitrary speakers in expressive VC has not been extensively explored. Previous approaches have relied on vocoders for speech reconstruction, which makes speech quality heavily dependent on the performance of vocoders. A major challenge of expressive VC lies in emotion prosody modeling. To address these challenges, this paper proposes a fully end-to-end expressive VC framework based on a conditional denoising diffusion probabilistic model (DDPM). We utilize speech units derived from self-supervised speech models as content conditioning, along with deep features extracted from speech emotion recognition and speaker verification systems to model emotional style and speaker identity. Objective and subjective evaluations show the effectiveness of our framework. Codes and samples are publicly available.\blfootnote{\textbf{Codes \& Speech Samples:} https://a2023aa.github.io/DEVC/ \\ Junchen Lu and Kun Zhou contributed to this work during their internship at UT Dallas.} 
\end{abstract}

\section{Introduction}


\label{sec:intro}
Emotions play a vital role in natural speech, as they convey one’s feelings, moods, and personality~\cite{schuller2020review}. Speech expressiveness encompasses a diverse range of emotions \cite{kun-mix}.  
Expressive voice conversion aims to jointly perform speaker identity and style transfer for emotional speakers, which poses huge potential in movie dubbing, voice acting, and human-computer interaction ~\cite{du2021expressive,gan2022iqdubbing,du22c_interspeech}.   

Voice conversion (VC) is the task of converting speaker identity while preserving linguistic content \cite{9262021}. Early VC studies focus on learning a statistical mapping between speech recordings of source and target speakers \cite{toda2007voice,aihara2012gmm}. Recent advancements in deep learning-based approaches significantly improve the performance of VC by learning the disentanglement across various speech characteristics. For instance, separate encoders guided by appropriate constraints have been explored to learn disentangled content and speaker representations \cite{liu2018voice,qian2019autovc,wang2021vqmivc}. Several studies introduce information bottleneck \cite{qian2019autovc, tjandra2019vqvae,kobayashi2021crank,chen2021tvqvc}, instance normalization \cite{chen2021again,chou2019one} and mutual information \cite{wang2021vqmivc,10096399} to achieve better disentanglement between content and speaker components. We note that most VC studies primarily focus on neutral speech, overlooking the prosodic variations manifested in different emotions. Expressive VC aims to fill this gap.  
\begin{figure}[!tp]
    \centering
    \includegraphics[width=6.5cm]{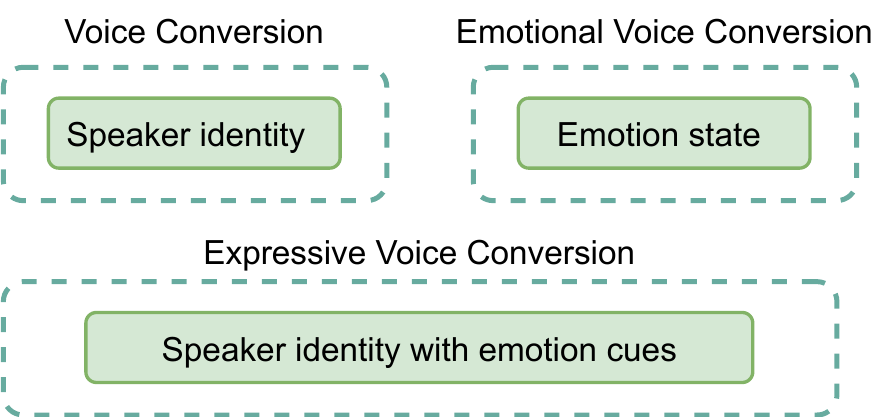}
    \caption{A comparison between the conversion of different speech components across VC\cite{9262021}, expressive VC\cite{du2021expressive}, and emotional VC\cite{zhou2021emotional}.}
    \label{fig:EVC}
\end{figure}

\begin{figure*}[t!]
    \centering
    \subfigure[Speaker representations with emotion cues ]{
    \begin{minipage}[b]{0.5\textwidth}
    \centering
    \includegraphics[width=8.3cm]{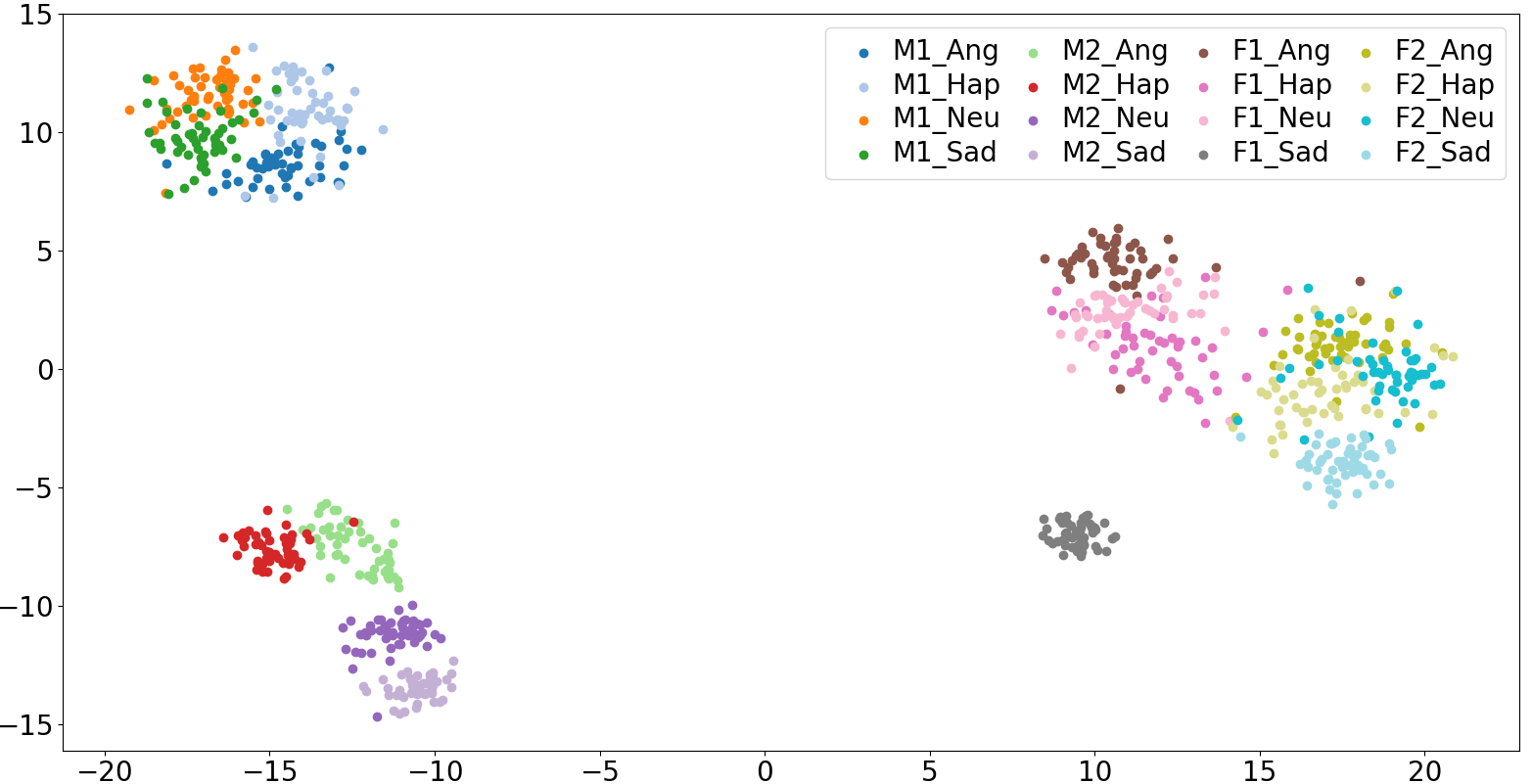}
    \label{spk_emb}
    \end{minipage}

    }\subfigure[Speaker-independent emotion representations]
    {
    \begin{minipage}[b]{0.5\textwidth}
    \centering
    \includegraphics[width=8.3cm]{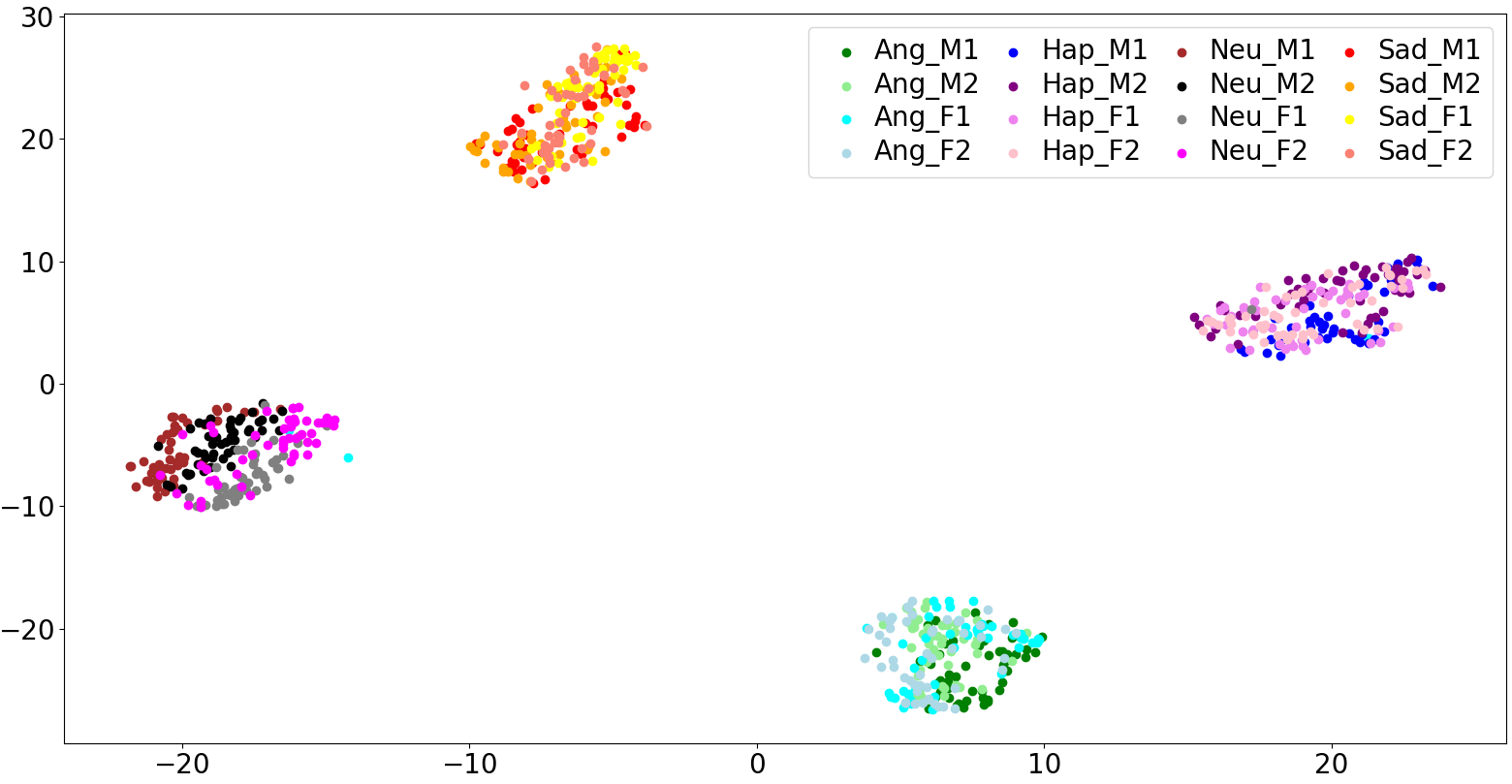}

    \label{emo_emb}
    \end{minipage}}  
    \caption{Visualization of speaker representations and speaker-independent emotion representations of 50 randomly selected utterances from 4 speakers across 4 different emotional states. Each point represents one expressive utterance and the legend indicates speaker identity and emotional state information.}
    \label{fig:spk_emb}
\end{figure*}

It remains a formidable undertaking to model emotional styles for expressive VC due to the hierarchical structure, inherent complexity, subjective nature, and variability of human emotional speech \cite{du2021expressive}. Moreover, emotional style contains both speaker-independent and speaker-dependent features \cite{sridhar2022unsupervised,zhou2021vaw, zhou2022emotion}.
Speaker-independent emotional features often exhibit consistent associations with the same emotion across different speakers \cite{zhou2020converting}, while speaker-dependent ones are manifested in speaker characteristics and are unique to each speaker \cite{sridhar2022unsupervised}.  
Despite their significant roles in conveying emotional style, speaker-dependent emotional features have been overlooked in previous VC studies \cite{du22c_interspeech,gan2022iqdubbing}. In this paper, we aim to enhance the expressiveness of converted speech for VC systems by exploring both speaker-dependent and speaker-independent emotional features.

Another limitation of previous approaches \cite{du22c_interspeech,gan2022iqdubbing,du2021expressive} of expressive voice conversion stems from their reliance on vocoders to reconstruct speech waveform from acoustic features. Consequently, the quality of the synthesized speech is intricately influenced by vocoder's performance~\cite{9747020}, which typically requires a substantial amount of high-quality speech data for training. To address these issues, we propose a fully end-to-end diffusion-based framework for expressive VC for the first time. We also demonstrate the flexibility of our model on any-to-any conversion.

In this paper, we introduce an expressive voice conversion framework, namely \textit{DEVC}, which enables effective speaker identity conversion for emotionally expressive speakers. To achieve this, we employ three encoders dedicated to handling content representation, speaker representation incorporating speaker-dependent emotional cues, and speaker-independent emotion representation, respectively. Content conditioning is facilitated through speech units, while deep features derived from speaker verification (SV) and speech emotion recognition (SER) tasks are utilized to capture speaker-dependent and speaker-independent emotional information. Furthermore, we employ a conditional denoising diffusion probabilistic model (DDPM)  to iteratively reconstruct the waveform from Gaussian noise, conditioned on the content, speaker, and emotion representations. Our main contributions include:

\begin{itemize}
    \item We propose a fully end-to-end expressive voice conversion framework based on a conditional diffusion model without the need for large-scale training data and manual annotations;
    \item Our findings reveal that speaker embeddings derived from an SV model pre-trained on neutral data effectively capture speaker-dependent emotional cues, thereby demonstrating their utility in enhancing expressive voice conversion; and
    \item Our proposed framework shows flexibility in identity conversion for both seen and unseen emotional speakers, achieving any-to-any expressive voice conversion. 
\end{itemize}

  
The rest of this paper is organized as follows: Section~\ref{sec:Related} provides an introduction to the related work. In Section~\ref{sec:Analysis}, we undertake a comprehensive analysis of the emotion and speaker representations employed in our framework. Section~\ref{sec:Main} presents the details of our proposed framework. The experimental results are reported in Section~\ref{sec:Experiments}. In Section~\ref{sec:Conclusion}, we conclude the study.

\section{Related Work}
\label{sec:Related}

\subsection{Expressive Voice Conversion}



Expressive speech is a complex composition of various components, including speaker identity, emotional style, and linguistic content. Previous models in expressive voice conversion \cite{du22c_interspeech, gan2022iqdubbing} primarily focus on eliminating redundant information between emotional style and other speech components, such as content and speaker identity. This objective has been achieved through techniques like mutual information loss \cite{du22c_interspeech} and prosody filters \cite{gan2022iqdubbing}. In another work \cite{du2021expressive}, a StarGAN-based model incorporates deep features from an SER model as emotion representation and a simple one-hot vector as the speaker representation. However, this simplistic speaker representation may not adequately capture the nuances of emotional speakers, thus limiting its ability to achieve any-to-any voice conversion for arbitrary speakers. 

Another related task is emotional voice conversion, which focuses on converting the emotional state of the speaker while preserving the speaker's identity \cite{10095842, zhou2021emotional}. In contrast, expressive voice conversion aims to simultaneously convert both speaker identity and speech style for speakers displaying emotional expression \cite{du2021expressive}, as shown in Figure \ref{fig:EVC}. This paper will concentrate on expressive voice conversion.


\subsection{Denoising Diffusion Probabilistic Models in Speech Synthesis}


Denoising diffusion probabilistic models \cite{ho2020denoising} are a class of diffusion models that uses a Markov chain to gradually convert a simple distribution into a complicated data distribution. It contains two processes: a diffusion process where clean input data is converted to an isotropic Gaussian distribution by adding noise step by step; and a reverse process where clean input can be recovered from Gaussian noise by predicting and removing the noise introduced in each step of the diffusion process. Denoising diffusion probabilistic models have achieved great success in various speech synthesis tasks, such as vocoding \cite{kong2021diffwave}, text-to-speech\cite{ijcai2022p577}, and VC\cite{liu2021diffsvc, popov2021diffusion, sadekova2022unified}. We further develop the idea and introduce an end-to-end framework for expressive voice conversion, which will be elaborated in Section \ref{sec:Main}. 


\subsection{Representation Learning with Self-Supervised Models}

Collecting labeled speech data is difficult, expensive, and time-consuming \cite{lin2021s2vc, mohamed2022self, chen22q_interspeech}. Self-supervised speech models pre-trained on a large amount of unlabeled data can learn high-level speech representations without relying on labeled data \cite{baevski2019vq, schneider2019wav2vec, baevski2020wav2vec, hsu2021hubert, yang23v_interspeech}.
Self-supervised representations have been proven to be highly effective in various speech synthesis tasks due to their ability to capture essential speech content information \cite{kreuk2022textless, lu2023high}.
The majority of existing VC methods leveraging such representations separate content from speaker-related information through discretization \cite{huang2021any, polyak21_interspeech}.
To retain more linguistic content in VC, van Niekerk et al. \cite{van2022comparison} propose to learn soft speech units by predicting a distribution over discretized representations, further improving the intelligibility of converted speech.
Inspired by this, we incorporate soft speech units as content conditioning in our proposed method. 

\begin{table*}[t!]
    \centering
    \caption{Mean Euclidean distance between speaker representations from different emotion pairs among 4 speakers. The first row and the first column respectively represent emotion states of two groups, with each cell indicating a unique emotion pair. Note that speakers here are the same set of speakers from Figure 2.}

    \subfigure[Speaker: 0013 (M1)]{

    \scalebox{0.9}{
    \begin{tabular}{c|cccc}
    \hline
          & Angry          & Happy          & Neutral        & Sad            \\ \hline
    Angry   & \textbf{0.670} & 0.724          & 0.761          & 0.739          \\
    Happy   & 0.722          & \textbf{0.719} & 0.773          & 0.754          \\
    Neutral & 0.753          & 0.765          & \textbf{0.676} & 0.703          \\
    Sad     & 0.729          & 0.751          & 0.705          & \textbf{0.667} \\ \hline
    \end{tabular}}
    }\hspace{1em}
\subfigure[Speaker: 0020 (M2)]{

\scalebox{0.9}{
\begin{tabular}{c|cccc}
\hline
       & Angry          & Happy          & Neutral        & Sad            \\ \hline
Angry   & \textbf{0.705} & 0.774          & 0.811          & 0.867          \\
Happy   & 0.784          & \textbf{0.669} & 0.761          & 0.783          \\
Neutral & 0.832          & 0.754          & \textbf{0.693} & 0.735          \\
Sad     & 0.873          & 0.776          & 0.733          & \textbf{0.615} \\ \hline
\end{tabular}}
}\hspace{1em}

\subfigure[Speaker: 0016 (F1)]{

\scalebox{0.9}{
\begin{tabular}{c|cccc}
\hline
        & Angry          & Happy          & Neutral        & Sad            \\ \hline
Angry   & \textbf{0.699} & 0.803          & 0.782          & 0.955          \\
Happy   & 0.811          & \textbf{0.751} & 0.783          & 0.904          \\
Neutral & 0.787          & 0.774          & \textbf{0.726} & 0.896          \\
Sad     & 0.952          & 0.896          & 0.890          & \textbf{0.682} \\ \hline
\end{tabular}}
}\hspace{1em} 
\subfigure[Speaker: 0018 (F2)]{
\centering
\scalebox{0.9}{
\begin{tabular}{c|cccc}

\hline
        & Angry          & Happy          & Neutral        & Sad            \\ \hline
Angry   & \textbf{0.672} & 0.745          & 0.725          & 0.808          \\
Happy   & 0.756          & \textbf{0.700} & 0.732          & 0.758          \\
Neutral & 0.724          & 0.733          & \textbf{0.668} & 0.780          \\
Sad     & 0.814          & 0.752          & 0.772          & \textbf{0.695} \\ \hline
\end{tabular}}
}
\label{fig:distance}
\end{table*}


\section{Emotion and Speaker Representations: A Novel Analysis for Expressive VC}
\label{sec:Analysis}

Expressive speech introduces variations in the acoustic features, resulting in increased complexity for speaker identity among emotional speakers \cite{parthasarathy2017study}. 
In expressive voice conversion, it is expected that speaker identity contains both speaker-dependent emotional cues and other speaker-related information. 
Speaker-dependent emotional cues refer to the emotional information that is specific to an individual speaker. These cues include variations in intonation, rhythm, and voice quality that reflect the speaker's emotional state. 

To incorporate such speaker-dependent emotional cues into our expressive voice conversion model, we extract speaker representations from a pre-trained SV model \cite{wan2018generalized}. These speaker representations serve as conditioning information for our expressive voice conversion model, allowing it to generate converted speech that preserves both target speaker identity and their emotional nuances.

By employing the t-SNE algorithm \cite{van2008visualizing}, we visualize the speaker representations of two female speakers and two male speakers, as shown in Fig. \ref{spk_emb}. We observe that the speaker representations form separate clusters for each speaker, indicating successful differentiation of speaker identities. We also notice the formation of smaller clusters within the same speaker and emotional state, indicating the presence of speaker-dependent emotional information. 

To further evaluate the similarity of speaker representations, we perform a comprehensive analysis. We randomly selected 240 emotional speech utterances from the ESD dataset~\cite{zhou2021emotional} and divided them equally into two groups.  We calculate the Euclidean distance between utterance pairs from these groups and obtain the mean distance of emotion pairs. As presented in Table \ref{fig:distance}, the results consistently demonstrate that the distances within the same emotional state are lower compared to those between different emotions. This consistent pattern strongly supports our hypothesis that the speaker representations not only capture speaker-related information but also encompass speaker-dependent emotional cues. 

In our proposed method, we adopt deep features obtained from a pre-trained speaker-independent SER model \cite{8421023} as speaker-independent emotion representations, following the approach in \cite{zhou2021seen}. To visually analyze the speaker-independent emotion representations, we employ the t-SNE algorithm \cite{van2008visualizing} and present the results in Figure \ref{emo_emb}. The visualization demonstrates that the emotion representations derived from different emotions form well-separated clusters. Within each cluster, representations from the same emotion but different speakers tend to be mixed up. This observation confirms that our emotion representations encompass speaker-independent emotional style information, allowing for generalization across different individuals. 

Our analysis provides valuable insights into the nature of our emotion and speaker representations.  
By incorporating these representations, our framework enables a single DDPM to synthesize expressive speech for any given speaker by effectively modeling both common characteristics of each emotional state across different speakers and emotional nuances associated with each individual. Further details on this approach will be provided in Section 4.


\begin{figure*}[t]
    \centering
    \includegraphics[width=16cm]{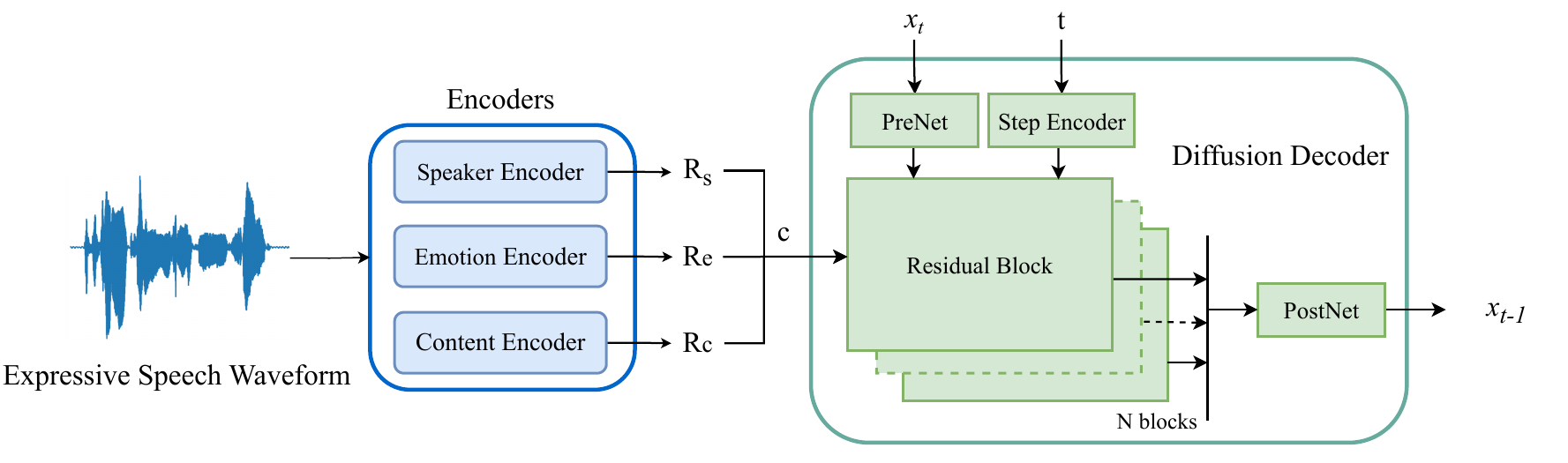}
    \caption {An illustration of the training phase of the proposed DEVC, where the green boxes represent the modules that are involved in the training while the others are not.} 
    \label{fig:train}
\end{figure*}

\section{Proposed Method - DEVC}
\label{sec:Main}

In this section, we introduce our fully end-to-end expressive VC model, Diffusion-based Expressive Voice Conversion (DEVC). 
DEVC consists of a content encoder, a speaker encoder, a speaker-independent emotion encoder, and a diffusion-based decoder, as illustrated in Figure \ref{fig:train}. These three encoders extract content representations, speaker representation with emotional cues, and speaker-independent emotion representations from input expressive speech, respectively. Utilizing these auxiliary representations as conditions, the diffusion-based decoder iteratively generates the converted expressive speech, starting from Gaussian noise. 

\subsection{Training Stage}
Given an expressive speech utterance with $S$ audio segments, the content encoder extracts a 256-dimensional content representation $R_c \in \mathbbm{R}^{S \times 256}$ from the waveform. The speaker encoder extracts a $256$-dimension utterance-level speaker representation ${R_s}$ with emotional cues, as reported in Table \ref{fig:distance} and Figure \ref{spk_emb}. The emotion encoder extracts an utterance-level speaker-independent emotional style representation $R_e$ with 128 dimensions. To align with $R_c$, we upsample the two utterance-level $R_s$ and $R_e$ to $R'_s \in \mathbbm{R}^{S \times 256}$ and $R'_e \in \mathbbm{R}^{S \times 128}$ by repeating them, respectively. We formulate the conditioning $c$ by segment-wise concatenating $R_c$ with $R'_s$ and $R'_e$:
\begin{equation}
    c= {\rm concat}(R_c, R'_s, R'_e)
\end{equation}

The conditional DDPM-based decoder has two subprocesses: the diffusion process and the reverse process. Given a segment of waveform $x_0$, the diffusion process is the process of $x_0$ being gradually corrupted to Gaussian noise $x_T$ within finite $T$ steps. 
Assume $\{x_1,...,x_{T-1}\}$ is a sequence of latent variables where $x_t$ is transformed from $x_{t-1}$ by adding Gaussian noises at each timestep $t\in[1,T]$: 

\begin{equation}
    q(x_t|x_{t-1}) = \mathcal N(x_t; \sqrt{1-\beta_t}x_{t-1},\beta_tI) \\
\end{equation}
where $\{\beta_1,...\beta_T\}$ is a fixed variance schedule. Given clean data $x_0$, sampling of $x_t$ can be written in a closed form:

\begin{equation}
\begin{gathered}
    q(x_t|x_0) =\mathcal N(x_t; \sqrt{\bar{\alpha}_t}x_0,(1-\bar{\alpha_t})I)\\
    x_t =\sqrt{\bar{\alpha}_t}x_0 + \sqrt{1-\bar{\alpha}_t}\epsilon
\end{gathered}
\end{equation}
where $\alpha_t=1-\beta_t$ and $\bar{\alpha_t}=\prod_{s=1}^{t}\alpha_s$. Noise $\epsilon\sim\mathcal N(0,I)$ has the same dimensionality as data $x_0$ and latent variables $x_1,...x_T$.



The reverse process generates a reverse sequence by
sampling the posteriors $q(x_{t-1}|x_t)$, starting from a Gaussian noise sample $x_T$. However, since
$q(x_{t-1}|x_t)$ is intractable, the decoder learn parameterized
Gaussian transitions $p_\theta(x_{t-1}|x_t)$ with a learned mean
$\mu_\theta(x_t,t,c)$ and a fixed variance $\sigma^2_tI$\cite{ho2020denoising}:
\begin{equation}
p_\theta(x_{t-1}|x_t)=\mathcal N(\mu_\theta(x_t,t,c),\sigma^2_tI)
\end{equation}
where $\mu_\theta(x_t,t,c)$ is the function of a noise approximator $\epsilon_\theta(x_t,t,c)$. 

Based on speech-conditioning pairs, we then learn the conditional diffusion-based decoder via:

\begin{equation}
L_\theta =\Vert \epsilon - \epsilon_\theta(x_t,t,c)\Vert^2_2
\end{equation}
where $\epsilon\sim\mathcal N(0,I)$ is the noise and $\epsilon_\theta$ denotes the decoder with learnable parameters $\theta$.

\subsection{Run-time Conversion}
\begin{figure}[!tp]
    \centering
    \includegraphics[width=6cm]{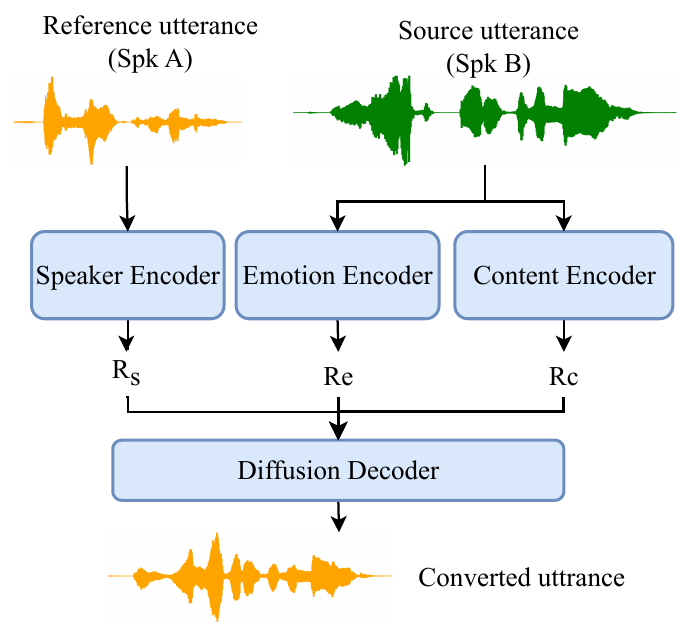}
    \caption{An illustration of the run-time phase of the proposed DEVC, highlighting that waveform generation occurs without the use of a vocoder.}
    \label{fig:run-time}
\end{figure}

At run-time, DEVC takes a source utterance from one speaker and a reference utterance from a target speaker, which is either seen or unseen during training, as shown in Figure \ref{fig:run-time}. The encoders extract content representation and speaker-independent emotional style representation from the source utterance and speaker representation with emotional cues from the reference utterance and the diffusion-based decoder generates converted expressive speech from Gaussian noise and timestep conditioned on those representations. We note that emotion representations are derived from a speaker-independent emotion recognizer, thereby inherently assumed to be speaker-independent (as shown in Section 3). Through our experiments, we illustrate that the emotion encoder can take input from either the source utterance or the reference.

\section{Experiments}


\begin{table*}[]
\caption{Objective evaluations results for the conversion between speakers seen during training (S2S), the conversion for seen to unseen speakers (S2U), and the conversion between unseen speakers (U2U). MCD is measured in dB.}
\scalebox{0.77}{
\begin{tabular}{cc|cccc|cccc|cccc|cccc}
\hline
\multicolumn{2}{c|}{{}}                                                                   & \multicolumn{4}{c|}{{Neutral}}                                                                      & \multicolumn{4}{c|}{{Angry}}                                                                        & \multicolumn{4}{c|}{{Happy}}                                                                        & \multicolumn{4}{c}{{Sad}}                                                                           \\ \cline{3-18} 
\multicolumn{2}{c|}{\multirow{-2}{*}{{}}}                                                 &  {MCD$\downarrow$}  &  {SV$\uparrow$}      &  {FFE$\downarrow$}  &  {VDE$\downarrow$}  &  {MCD$\downarrow$}  &  {SV$\uparrow$}      &  {FFE$\downarrow$}  &  {VDE$\downarrow$}  &  {MCD$\downarrow$}  &  {SV$\uparrow$}      &  {FFE$\downarrow$}  &  {VDE$\downarrow$}  &  {MCD$\downarrow$}  &  {SV$\uparrow$}      &  {FFE$\downarrow$}  &  {VDE$\downarrow$}  \\ \hline
\multicolumn{1}{c|}{{}}                      &  {Baseline\cite{du2021expressive}}   &  {8.86} &  {0.72} &  {0.39} &  {0.34} &  {8.79} &  {0.75} &  {0.39} &  {0.34} &  {8.84} &  {0.72} &  {0.47} &  {0.37} &  {9.08} &  {0.73} &  {0.41} &  {0.34} \\
\multicolumn{1}{c|}{\multirow{-2}{*}{{S2S}}} &  {\textbf{DEVC}}                &  {8.18} &  {0.84} &  {0.32} &  {0.29} &  {7.99} &  {0.87} &  {0.33} &  {0.30} &  {8.42} &  {0.82} &  {0.41} &  {0.34} &  {8.01} &  {0.80} &  {0.34} &  {0.30} \\ \hline
\multicolumn{1}{c|}{{}}                      &  {Baseline-U} &  {8.81} &  {0.68}        &  {0.42} &  {0.37} &  {8.74} &  {0.53}        &  {0.46} &  {0.39} &  {8.54} &  {0.71}        &  {0.47} &  {0.39} &  {8.51} &  {0.67}        &  {0.43} &  {0.35} \\
\multicolumn{1}{c|}{\multirow{-2}{*}{{S2U}}} &  {\textbf{DEVC}}                &  {8.80} &  {0.80}        &  {0.34} &  {0.30} &  {8.61} &  {0.71}        &  {0.39} &  {0.34} &  {8.43} &  {0.73}        &  {0.42} &  {0.36} &  {8.12} &  {0.76}        &  {0.39} &  {0.33} \\ \hline
\multicolumn{1}{c|}{{}}                      &  {Baseline-U} &  {9.02} &  {0.65}        &  {0.42} &  {0.37} &  {9.26} &  {0.69}        &  {0.47} &  {0.40} &  {8.72} &  {0.70}        &  {0.48} &  {0.39} &  {8.62} &  {0.67}        &  {0.45} &  {0.38} \\
\multicolumn{1}{c|}{\multirow{-2}{*}{{U2U}}} &  {\textbf{DEVC}}                &  {8.67} &  {0.79}        &  {0.35} &  {0.31} &  {8.62} &  {0.84}        &  {0.38} &  {0.33} &  {8.39} &  {0.79}        &  {0.41} &  {0.35} &  {8.19} &  {0.77}        &  {0.40} &  {0.35} \\ \hline
\end{tabular}}
\label{obj}
\end{table*}

\label{sec:Experiments}


\subsection{Dataset}

We evaluate the proposed DEVC on the ESD dataset\cite{zhou2021emotional}. For each emotion and each speaker, there are 300 training utterances, 20 reference utterances, and 30 test utterances. In our experiments, we select four emotions (Neutral, Angry, Happy, and Sad) and four speakers (0013, 0016, 0018, and 0020) as seen speakers for training, while randomly selecting four other speakers as unseen speakers. We evaluate our model in three scenarios: the conversion between seen speakers (S2S), the conversion from seen speakers to unseen speakers (S2U), and the conversion between unseen speakers (U2U), commonly referred to as the any-to-any task. All speech samples are sampled at 16 kHz and saved in 16-bit format.

\subsection{Experimental Setup}

DEVC consists of a content encoder, a speaker encoder, an emotion encoder, and a diffusion-based decoder. The content encoder is based on the HuBERT-Base backbone network, augmented with a linear layer \cite{van2022comparison} and pre-trained on the LibriSpeech-960 dataset \cite{7178964}. The emotion encoder consists of a three-dimensional CNN layer, a BLSTM layer, an attention layer, and a fully-connected (FC) layer. It is initially pre-trained on the IEMOCAP dataset \cite{busso2008iemocap} and subsequently fine-tuned using the ESD dataset \cite{zhou2021seen}. The speaker encoder follows the architecture in \cite{wan2018generalized}, utilizing a 3-layer LSTM with projection to generate 256-dimensional representations.

For the diffusion decoder, the architecture is similar to that described in \cite{kong2021diffwave}. The step encoder comprises a 128-dimensional sinusoidal position encoding and two FC layers with Switch activations. The PreNet consists of a Conv1$\times$1 layer followed by ReLU activation. The output of the step encoder and latent variables are then fed into a stack of 64 residual blocks with 128 residual channels. The skip connections from all residual layers are summed before entering the PostNet, which includes two Conv1$\times$1 layers. DEVC is trained on a single Nvidia 3080Ti GPU for 1.2 million steps using Adam optimizer with a batch size of 16 and a learning rate of 0.0002. 

\subsection{Baselines}
In a comparative study, we adopt the following two expressive VC models as our baseline:
\begin{itemize}
    \item \textbf{Baseline}: JES-StarGAN \cite{du2021expressive}, a many-to-many expressive voice conversion framework for S2S scenarios; and
    \item \textbf{Baseline-U}: We replace the one-hot speaker label in JES-StarGAN \cite{du2021expressive} with speaker representations for S2U and U2U settings. 
\end{itemize}

\begin{table}[t]
\centering
\caption {The MOS results with 95\% confidence interval to evaluate the speech quality of converted speech. }
\scalebox{1}{
\begin{tabular}{c|ccc}
\hline
Framework      & S2S & S2U & U2U \\ \hline
Baseline\cite{du2021expressive}          & 2.12$ \pm $0.15          & NA     & NA  \\ 
Baseline-U            & NA          & 2.45$ \pm $0.16       &2.46$ \pm $0.18  \\ \hline
\textbf{DEVC}           & 3.28$ \pm $0.08          & 3.16$ \pm $0.12       & 3.27$ \pm $0.13\\ \hline
\end{tabular}}
\label{tab:mos}
\end{table}

\begin{table}[t]
\centering

\caption {Ablation studies results in S2S setting.} 

\scalebox{1}{
\begin{tabular}{c|c|c|c}
\hline
Model      & MCD [dB] $\downarrow$ & SV $\uparrow$ & F0-RMSE [Hz] $\downarrow$ \\ \hline
$R_{c}$, $R_{s}$             &10.81           & 0.71      & 82.38 \\ 
$R_{c}$, $R_{e}$            &8.34          &  0.79      & 55.62\\ \hline
 $\boldsymbol{R_{c}}$, $\boldsymbol{R_{e}}$, $\boldsymbol{R_{s}}$ & \textbf{8.15}  & \textbf{0.85}  &\textbf{53.98}\\ \hline
\end{tabular}}
\label{tab:ab}
\end{table}

\subsection{Objective Evaluations}

Table \ref{obj} presents the objective evaluation results of our proposed framework and the baselines. Mel-cepstral distortion (MCD)\cite{kubichek1993mel} is utilized to measure spectral distortion between synthesized samples and target samples. Lower MCD values indicate better quality and closer similarity between the synthesized samples and the target samples \cite{9306487}. Our proposed method outperforms the baselines in all settings (S2S, S2U, and U2U) by achieving consistently lower MCD values, which indicates a better conversion performance. 

To evaluate speaker similarity, we utilize a pre-trained SV model\footnote{https://github.com/resemble-ai/Resemblyzer} and report the SV accuracy in Table \ref{obj}. Higher SV accuracy indicates better conversion of speaker identity \cite{lin2021s2vc}. Our proposed framework achieves higher speaker similarity compared to the baselines, highlighting its capability to accurately convert the characteristics of emotional speakers.

In addition, we employ two pitch similarity metrics: Voicing Decision Error (VDE) \cite{nakatani2008method} and F0 Frame Error (FFE) \cite{chu2009reducing}. VDE measures the proportion of voiced/unvoiced decision difference between the converted and target utterances, while FFE captures both pitch similarity and voiced/unvoiced decision differences \cite{polyak21_interspeech}. Our proposed method outperforms the baselines, indicating its effectiveness in capturing and reproducing the desired characteristics in the converted speech. 


\subsection{Subjective Evaluation}


We conduct listening experiments with 13 subjects to assess the speech quality, speaker similarity, and emotional style similarity. Each of them listens to 192 converted utterances in total. In the Mean Opinion Score (MOS) test, participants evaluated the converted utterances using a rating scale ranging from 1 (poor) to 5 (excellent) to assess speech quality. The results for S2S, S2U, and U2U are presented in Table \ref{tab:mos}. As shown in Table \ref{tab:mos}, DEVC outperforms the baselines in terms of speech quality. 

\begin{figure}[t]
    \centering
    \subfigure[Speaker Similarity]{     \begin{minipage}[b]{0.5\textwidth}
    \centering
    \includegraphics[width=8cm]{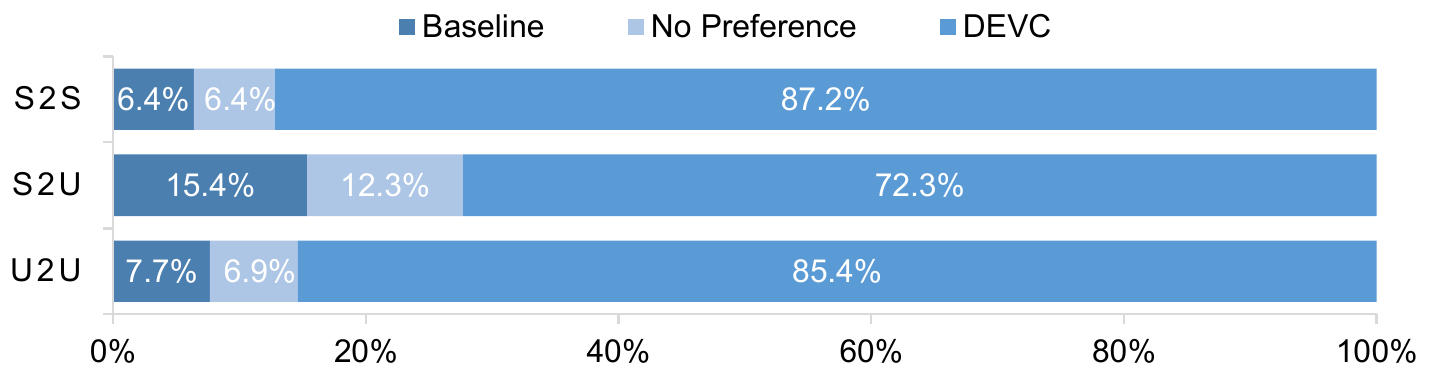}
    \label{spk_xab}
    \end{minipage}
    } 
    \subfigure[Emotional Style Similarity ]{  \begin{minipage}[b]{0.5\textwidth}
    \centering
    \includegraphics[width=8cm]{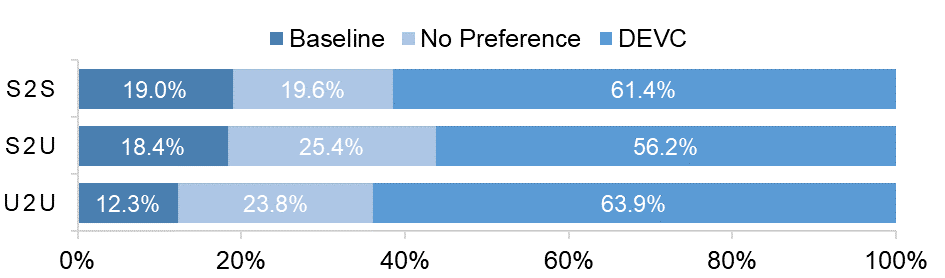}
    \label{emo_xab}

    \end{minipage}
    } 
    \caption{ABX preference results for S2S, S2U, and U2S settings to evaluate: (a) speaker similarity and (b) emotional style similarity. We used Baseline (JES-StarGAN) for S2S, and Baseline-U for S2U and U2S settings.}
   
    \label{xab}
\end{figure}

In addition, we perform ABX preference tests to evaluate speaker similarity in the S2S, S2U, and U2U settings. The listeners compare the converted utterances and select the utterance that sounds closer to the reference in terms of speaker identity. The results, illustrated in Figure \ref{spk_xab}, demonstrate that our proposed framework significantly outperforms the baseline, indicating that our method excels in accurately converting speaker identities for both seen and unseen speakers. 

We further conduct ABX preference tests to evaluate the emotional style similarity in S2S, S2U, and U2U settings. These tests require listeners to compare the emotional styles of the reference and converted utterances. Figure \ref{emo_xab} presents the results, clearly demonstrating the superior performance of our proposed framework over the baseline. These findings highlight the capability of our method to successfully convert emotional styles for arbitrary speakers, further validating the effectiveness of our emotional style modeling approach.


It is worth mentioning that the conversion between unseen speakers to unseen speakers is particularly challenging due to the lack of training data. However, even in this challenging scenario, our proposed method achieves excellent results in subjective evaluation, further affirming its ability to conduct high-quality conversion in any-to-any setting.

\subsection{Ablation studies}

We conduct ablation studies to analyze the impact of our speaker representations with emotion cues and speaker-independent emotion representations in the context of expressive voice conversion. Specifically, we investigate three different models,  all sharing the same diffusion-based decoder architecture: the first model includes content representations and speaker representations, denoted as $R_c, R_s$. The second model incorporates content representations, speaker-independent emotional style representations, and one-hot speaker labels, denoted as $R_c, R_e$. The third model represents our proposed method, incorporating content representations, speaker representations with emotion cues, and speaker-independent emotion representations, denoted as $R_c, R_e, R_s$. 

In the S2S setting, we evaluate the performance of these models by calculating metrics MCD, SV accuracy, and F0 root mean squared errors (F0-RMSE)\cite{luo2019neutral}, as presented in Table \ref{tab:ab}. 
From the results, our proposed method, which utilizes speaker representations instead of one-hot speaker labels, exhibits superior performance compared to the model incorporating one-hot speaker vectors. This highlights the advantage of our speaker representation with emotion cues, indicating that it captures and represents speaker-dependent emotion cues more effectively. In addition, we observe that our proposed method outperforms the other models. This finding suggests the effectiveness of our emotional style modeling method in the expressive voice conversion task.

\section{Discussion}
We note that our proposed method has the potential to be extended to convert emotion states for different speakers with various reference utterance settings. However, in this paper, our focus is solely on expressive VC. We intend to broaden the scope of this work to include the conversion of emotion states for the same speaker (emotional VC) or different speakers in our future research.

\section{Conclusion}
\label{sec:Conclusion}

In this paper, we present a novel framework for expressive voice conversion using a diffusion-based approach that supports any-to-any conversion. Our framework leverages speech units as content representations and incorporates deep features extracted from speech emotion recognition and speaker verification tasks to capture emotion and speaker characteristics. A key finding of our research is that speaker embeddings obtained from a pre-trained SV model using neutral data inherently contain speaker-dependent emotional features, thereby proving advantageous for the expressive voice conversion task. Our proposed framework exhibits remarkable flexibility in converting both seen and unseen speakers without the need for vocoders, marking it as the first end-to-end diffusion model-based expressive voice conversion framework known to us. 
\newpage
\footnotesize
\bibliographystyle{IEEEbib}
\bibliography{Odyssey2024_BibEntries,refs}

%

\end{document}